\documentstyle[12pt,aaspp4,flushrt]{article}
\doublespace
\begin{document}
\title{ Gamma-Ray Bursts: Afterglows and Central Engines 
}
\author{K.S. Cheng $^1$, T. Lu$^2$}
\affil{
$^1$Department of Physics, University of Hong Kong, Pokfulam Road, Hong 
Kong, China\\
$^2$ Department of Astronomy, Nanjing University, Nanjing 210093, 
China}

\begin{abstract} 

Gamma-ray bursts (GRBs) are most intense transient gamma-ray events in the sky when they are on together with the strong evidences (i.e. the isotropic and inhomogeneous distribution of GRBs detected by BASTE) that they are located at cosmological distances, which make them the most energetic events ever known. For example, the observed radiation energies of some GRBs are equivalent to convert more than one solar mass energy into radiation completely. This is thousand times stronger than that of supernova explosion. Unconventional energy mechanisms and extremely high conversion efficiency for these mysterious events are required.  The discovery of host galaxies and association with supernovae in the cosmological distances by the recently launched satellite of BeppoSAX and ground based radio and optical telescopes in GRB afterglow provides further support to the cosmological origin of GRBs and put strong constraints on central engines of GRBs. It is the aim of this article to review the possible central engines, energy mechanisms, dynamical and spectral evolution of GRBs, especially focusing on the afterglows in multi-wavebands.

\end{abstract} 

\newpage

\section{Introduction}

The nature of gamma-ray burst is still a mystery since its discovery 
about thirty years ago (Klebesadel, Strong \& Olsen 1973). It has been thought that  Galactic magnetized neutron 
stars are promising candidates to the gamma-ray burst sources because 
of the spectral features that had been observed in several different early instruments (Mazet et al. 1981; Hunter 1984; Fenimore et al. 1988; Murakami et al. 1988). However, the launch of Compton Gamma-Ray 
Observatory (CGRO) on April 5, 1991 has led to a new insight to these 
gamma-ray sources. From the spatial distribution of the gamma-ray bursts 
observed by the Burst and Transient Source Experiment (BATSE) on board CGRO, it was found that they are more consistent to  
isotropic distribution, which is difficult to be understood by the galactic 
neutron star models. In addition, the smaller number of weak sources observed 
in comparing to that of strong sources can be interpreted as we are 
observing the 
boundary of the sources. Many models then have been suggested that 
these high energy gamma-ray sources may originate from extragalactic region. 
The isotropy of the sources can then be explained and the deficit of weak 
sources can be understood.
However,  most of these cosmological 
models fail to give a quantitative explanation to  the spectral features (lines and continuum) observed 
in the gamma-ray bursts.

In early 1997, the Italian-Dutch Satellite, BeppoSAX, brought the great
break-through by rapid and accurate GRB localization and thus provided a 
few arcmins (or even smaller) error box. With so small an error box, 
it identified an X-ray counter-part (now known as X-ray afterglow) 
of GRB 970228 just 8 hours after the $\gamma$-ray trigger (Costa et al. 1997). 
Several hours later, its optical afterglow was also observed 
(Groot et al. 1997; van Paradijs et al. 1997).
Since then, BeppoSAX has observed more than 20 GRBs, of which almost all 
exhibited X-ray afterglows. Based on the precise localization, many 
telescopes observed about a dozen optical afterglows and about ten radio 
afterglows. Up to now, people have observed host galaxies of more than 
ten GRBs with large red-shifts, showing them definitely at cosmological 
distances. These great discoveries lead to rapid developments in GRB studies(van 
Paradijs et al. 2000). A lot of 
questions have now been clarified. However, compared with GRB itself, afterglow 
appears to be simpler and has been known much better. In contrast, the GRB 
itself, especially its energy source and origin, still keeps to be mysterious.

The main observational facts of GRBs are as follows:
                                                                                                         
\hspace*{1cm} {\bf(i)Burst Rate:}

BATSE has detected about one GRB per day from an unpredictable direction in the sky. It is believed that the burst rate could be up to $10^3$ bursts per year if the actual solid angle and the sky exposure time are corrected.

\hspace*{1cm} {\bf(ii)Temporal Properties:}

 The majority of GBRs has a very complex temporal structures. Their variability $\delta T$ is significantly shorter than the duration $T$. Typically, $\delta T \sim 10^{-2} T$. $T$ is very short, usually only a few seconds to tens of seconds, but occasionally it could be as long as a few tens of minutes
or as short as a few milli-seconds.  There seems to be a roughly 
bimodal distribution of long bursts of $T \geq 2$ s and short bursts 
of $T \leq 2$ s (Kouveliotou et al. 1993). The time scales of variability ($\delta T$), especially their rising time, may be as short as only milli-seconds or even sub-milli-seconds. There are various time dilations in temporal profiles of GRBs. If the GBRs are cosmological origin then the temporal profiles and spectra of more distant sources will be time dilated compared to those relatively nearby sources (e.g. Norris et al. 1994; Wei \& Lu 1997). Time dilations in different energy channels were also discovered (e.g. Cheng et al. 1995; 1996b).

\hspace*{1cm} {\bf (iii)Spatial distribution:} 

The angular distribution of GRBs' position on the sky is perfectly isotropic
(Greiner 1999). For the first 1005 BATSE bursts the observed dipole
and quadrupole (corrected to BATSE sky exposure) relative to the galaxy are
0.017 $\pm$ 0.018 and  $-0.003 \pm$ 0.009 respectively. These values are 0.9$\sigma$ and 0.3$\sigma$ deviated from complete isotropy (Meegan et al. 1992; Briggs 1995) respectively. This distribution favors GRBs at cosmological distances at least statistically. However, GRBs at extended dark halo of our Galaxy could also explain this feature. This led to a great debate between galactic origin and cosmological origin. There are a few strong bursts and many more weak bursts. A sample of the first 601 bursts is used to analyze the distribution of sources in space and shows a 14$\sigma$ deviated from the homogeneous flat space distribution (Pendleton et al.1995) but compatible with a cosmological distribution. However, the distribution of short bursts is not inconsistent with homogeneous Euclidian distribution (Katz \& Canel 1996).

\hspace*{1cm} {\bf (iv)Total Energy of the Burst:}
The $\gamma$-ray fluences are typically in the range of $(0.1 - 10) \times 10^{-6}$ ergs/cm$^{2}$. If the possible beaming is ignored, these fluences imply that the total burst energies are $10^{45}-10^{46}$ergs for galactic halo distance and $10^{52}-10^{53}$ergs for cosmological distance. However, the energy release of GRB 990123 in $\gamma$-rays is already as large as about $3.4 \times 10^{54}$ ergs, if considering the low efficiency of $\gamma$-rays radiation, the total energy release could be a few times of $10^{55}$ ergs or even much larger, this imposes strong challenge for all GRB models(Kulkarni et al. 1999a; Andersen et al. 1999).

\hspace*{1cm} {\bf (v)Spectral properties:}

Most of GRB power is radiated in the 100 - 1000 keV range, but photons up to
18 GeV or down to a few keV have also been registered. The non-thermal spectra
of GRBs are best fitted by a broken power-law with two
spectral indices. If a simple power-law fit is used,
namely F($E_\gamma$) $\propto$ $E_{\gamma}^{-\alpha}$, where $\alpha$
is about 1.8 --- 2 (Piran 1999). The low energy part of
the spectrum behaves in many cases like a simple power-law:
F $\propto E_{\gamma}^{-\beta}$ with $-1/2 < \beta < 1/3$ (Katz 1994; Cohen et al. 1997). This can be easily explained in terms of the non-uniform magnetic distribution inside the fireball shock front (Cheng \& Wei 1996). Observations by earlier detectors as well as by BATSE have shown that the spectrum varies during the bursts.   Most bursts evolve from hard to soft, but different trends were found (Norris et al.  1986; Ford et al. 1995). Both absorption and emission features had been reported  by various experiments prior to BATSE. Absorption lines in the 20 - 40 keV range had been observed by several experiments (Murakami et al. 1988; Fenimore et al.1988), but never simultaneously. Emission features near 400 keV had been claimed in other bursts (Mazets et al. 1980). However BATSE has not found any of the spectral
features (Palmer et al. 1994; Band et al. 1996).

\hspace*{1cm} {\bf (vi)Afterglows:} 

Afterglows are counterparts of GRBs at wave bands other than 
$\gamma$-rays, may be in X-ray, optical, or even radio bands (Costa et al. 1997; van Paradijs et al. 1997;
Galama et al. 1997; Frail et al. 1997; Taylor et al. 1997; Piran 1999). They are variable, 
typically decaying according to power laws: $F_{\nu} \propto t^{-\alpha}$ 
($\nu=$ X, optical, ......) with $\alpha = 1.1 - 1.6$ for X-ray, $\alpha 
= 1.1 - 2.1$ for optical band. X-ray afterglows can last days or even weeks; 
optical afterglows and radio afterglows months. The most important discovery 
is that many 
afterglows show their host galaxies being definitely at cosmological distances 
(with large red-shifts up to $Z=3.4$ or even 5). Thus, the debate is 
settled down, GRBs are at cosmological distances, they should be the most 
energetic events ever known since the Big Bang.

\section{The Standard Fireball Shock Model}

\hspace*{1cm} {\bf (i)Stellar Level Event:} 

The variability time scale is usually very short. Let $\delta T \sim {\rm ms}$, then, the space scale of the initial source, 
$R_{\rm i} <c\delta T \sim 3 \times 10^{2}$ km. Hence, even for black hole, considering $R = 2{\rm G}M/c^{2}$, we have $M \leq c^{3} \delta T / (2 {\rm G}) \sim 10^{2}$~M$_{\odot}$. If the burster is not a black hole, its mass should be much smaller. Thus, we can conclude that the GRB should be 
{\bf a stellar phenomenon} and the burster should be {\bf a compact 
stellar object} which may be related with neutron star (or strange star) or 
stellar black hole. 

\hspace*{1cm} {\bf (ii)Fireball:} 

>From the measured fluence $F$ and the measured distance $D$, 
if emission is isotropic, we can calculate the total radiated energy to be $E_{0} = F(4\pi 
D^{2})\approx 10^{51}(F/(10^{-6}{\rm ergs/cm}^{2}))(D/(3{\rm Gpc}))^{2}$. Thus, 
very large energy ($10^{51}$ ergs) is initially contained in a small volume of 
$(4/3)\pi R_{i}^{3} \sim 1 \times 10^{23}$ cm$^{3}$. This should be inevitably 
a fireball, of which the optical depth for 
$\gamma \gamma \longrightarrow {\rm e}^{+} {\rm e}^{-}$, $\tau_{\gamma\gamma}$, 
is very large. Consider a typical burst
its average optical depth can be written as: 
\begin{equation}
\tau_{\gamma\gamma} = \frac{f_{\rm p} \sigma_{\rm T} F D^{2}}{R_{\rm i}^{2}
m_{\rm e} c^2}
\sim 10^{17} f_{\rm p}
\bigg(\frac{F}{10^{-6} {\rm ergs/cm^2}}\bigg)
\bigg(\frac{D}{3~{\rm Gpc}}\bigg)^2
\bigg(\frac{\delta T}{1~{\rm ms}}\bigg)^{-2},  
\end{equation}
where $f_{\rm p}$ denotes the fraction of photon pairs 
satisfying $\sqrt{E_1 E_2} > m_{\rm e} c^2$. 

For so large an optical depth, there seem to appear two serious difficulties. 
First, the radiation in an optically thick case should be thermal, while 
the observed radiation is definitely non-thermal. Second, high energy 
photons should be easily converted into ${\rm e}^{+} {\rm e}^{-}$ pairs, 
while the observed high energy tail indicates that 
this convertion has not happened. However, it is very interesting to note 
that just such a large optical depth paves the way to solve both of them.

\hspace*{1cm} {\bf (iii)Compactness Problem:} 

In fact, the luminosity of the thermal 
radiation, according to the 
Stefan-Boltzmann law, should be proportional to the surface of the 
fireball which is initially so small that the thermal radiation can not 
be observed. However, just due to the large optical depth, the radiation 
pressure should be very high and could accelerate the fireball expansion 
to become ultra-relativistic with a large Lorentz factor $\gamma$. After 
expanding to a large enough distance, it may be getting optically thin.
At this time, the non-thermal $\gamma$-ray bursts can be observed. Does 
such a large distance contradict the compactness relation 
$R_{\rm i} \le {\rm c} \delta T$ compared with the 
milli-second variabilities? To answer this question, let us first 
note that this relation holds only for non-relativistic (rest) 
object with $R_{\rm i}$ denoting its space scale. For an ultra-relativistically expanding fireball, the compactness relation should be relaxed to 
\begin{equation}
R_{\rm e} \le \gamma^{2} {\rm c} \delta T,
\end{equation}
here $R_{\rm e}$ is the space scale of the expanding fireball with Lorentz 
factor $\gamma$. Considering two photons we observed at two different times 
apart by $\delta T$, as the emitting region is moving towards the observer 
with a Lorentz factor $\gamma \gg 1$, the second photon
should be emitted at a far nearer place than the first one. This gives 
effectively short time variabilities and leads to the additional factor 
$\gamma^{2}$ appearing in the above compactness relation.

The factor $f_{\rm p}$ in the optical depth $\tau_{\gamma\gamma}$ also 
sensitively depends on the ultra-relativistic expansion of the fireball. 
As for this case, the observed photons are blue-shifted, in the comoving 
frame, their energy 
should be lower by a factor of $\gamma$, and fewer photons will have 
sufficient energy to produce pairs. This gives a factor depending 
on spectral index $\alpha$, namely a factor of $\gamma^{2\alpha}$ in 
$\tau_{\gamma\gamma}$. 

\hspace*{1cm} {\bf (iv)Ultra-relativistic expansion:} 

Therefore, the optical 
depth $\tau_{\gamma\gamma}$ will decrease by a 
factor of $\gamma^{4+2\alpha}$ 
for the ultra-relativistically expanding fireball (Goodman 1986; 
Paczy\'{n}ski 1986; Piran 1999; Krolik \& Pier 1991):
\begin{equation}
\tau_{\gamma\gamma} =
{f_{\rm p} \over \gamma ^{2 \alpha}} {{\sigma_{\rm T} F D^2} 
  \over {R_{\rm e}^2 {\rm m}_{\rm e} {\rm c}^2}}
\approx {10^{17} \over \gamma ^{(4+2\alpha)}}
f_{\rm p}
\bigg({ F \over 10^{-6} {\rm ergs/cm^2}} \bigg)
\bigg({ D \over 3~{\rm Gpc}}\bigg)^2
\bigg({ \delta T \over 1~{\rm ms}} \bigg)^{-2}.
\end{equation}
Note, the spectral index $\alpha$ is approximately 2, we will have 
$\tau_{\gamma\gamma} < 1$ for $\gamma > 10^{17/(4+2\alpha)} \sim 10^{2}$. 
Thus, in order for the fireball to become optically thin, as required by 
the observed non-thermal spectra of $\gamma$-ray bursts, its expanding 
speed should be ultra-relativistic with Lorentz factor 
\[\gamma > \sim 10^{2}.\]
This is a very important character for GRBs, which 
limits the baryonic mass contained in the fireball seriously. If the initial
energy is $E_{0}$, then the baryonic mass $M$ should be less than 
\begin{equation}
E_{0}/({\rm c}^{2}\gamma) \approx 10^{-5}{\rm M}_{\odot} (E_{0} / (2\times 10^{51} {\rm ergs})),
\end{equation}
otherwise, the initial energy can not be converted to the kinetic energy of 
the bulk motion of baryons with such a high Lorentz factor. Most models related 
with neutron stars contain baryonic mass much higher than this limit. This is 
the famous problem named as ``baryon contamination''.

It is worthwhile to note that this very condition $\gamma > \sim 10^{2}$ can
also explain the existence of the high energy tail in the GRB spectra, as the 
observed high energy
photons should be only low energy photons in the frame of emitting region, they
are not energetic enough to be converted into ${\rm e}^{+}{\rm e}^{-}$ pairs.

\hspace*{1cm} {\bf (v)Internal-external shock:}

What is the radiation mechanism in the fireball 
model? The fireball 
expansion has successfully made a conversion of the initial internal energy 
into the bulk kinetic energy of the expanding ejecta. However, this is the 
kinetic energy of the associated protons, not the photons. We should have 
another mechanism to produce radiation, otherwise, even after the fireball 
becoming optically thin, the $\gamma$-ray bursts can not be observed. 
Fortunately, the shocks described below can do such a job. 

The fireball can be regarded as roughly homogeneous in its local rest frame, 
but due to the Lorentz contraction, it looks like a shell (ejecta) with 
width of the initial size of the fireball. As the shell collides with 
inter-stellar medium (ISM), shocks will be 
produced (Rees \& M\'{e}sz\'{a}ros 1992; Katz, J., 1994; 
Sari \& Piran 1995; Mitra 1998). These are usually 
called as external shocks. Relativistic electrons that have been 
accelerated in the relativistic shocks will usually emit synchrotron 
radiation. As the amount of swept-up interstellar matter getting larger 
and larger, the shell will be decelerated and radiation of longer wave 
length will be emitted. Thus, an external shock can produce only smoothly 
varying time-dependent emission, not the spiky multi-peaked structure 
found in many GRBs. If the central energy source is not completely 
impulsive, but works intermittently, it can produce many shells (or 
many fireballs) with different Lorentz factors. Late but faster shells 
can catch up and collide with early slower ones, and then, shocks 
(internal shocks) thus produced will lead to the observed bursting 
$\gamma$-ray emission (Rees \& M\'{e}sz\'{a}ros 1994; 
Paczy\'{n}ski \& Xu 1994). This 
is the so called internal-external shock model, 
internal shocks give rise to $\gamma$-ray bursts and external shocks to 
afterglows. The internal shocks can only convert a part of their energies 
to the $\gamma$-ray bursts, other part remains later to interact with the 
interstellar medium and lead to afterglows. Typically, the GRB is produced 
at a large distance of about 10$^{13}$ cm to the center, such a large 
distance is allowed according to the relaxed compactness relation 
$R_{\rm e} \le \gamma^{2} {\rm c} \delta T$, while its afterglows are produced at 
about 10$^{16}$ cm or even much farther. This internal-external shock 
scenario, under the simplified assumptions of uniform 
environment with typical ISM number density of $n \sim 1  {\rm cm^{-3}}$, 
isotropic emission of synchrotron radiation and only impulsive energy 
injection, is known as the standard model.  

\hspace*{1cm} {\bf (vi)Spectra of afterglows:} 

The instantaneous spectra of afterglows, according 
to this model, can be 
written as $F_{\nu} \propto \nu^{\beta}$, with different $\beta$ for 
different range of frequency $\nu$ (Sari et al. 1998; Piran 2000). Let 
$\nu_{sa}$ be the self absorption frequency, for which the optical depth 
$\tau(\nu_{sa})=1$. For $\nu < \nu_{sa}$, we have the Wien's law: 
$\beta=2$. For $\nu_{sa} < \nu < \min~(\nu_{\rm m},\nu_{\rm c})$, we 
can use the low energy synchrotron tail, $\beta=-1/3$. Here 
$\nu_{\rm m}$ is the synchrotron frequency of an electron with 
characteristic energy, $\nu_{\rm c}$ is the cooling frequency, namely 
the synchrotron frequency of an electron that cools during the local 
hydrodynamic time scale. For frequency within $\nu_{\rm m}$ and 
$\nu_{\rm c}$, we have $\beta = -1/2$ for fast cooling 
($\nu_{\rm c} < \nu_{\rm m}$) and $\beta=-(p-1)/2$ for slow cooling 
($\nu_{\rm m} < \nu_{\rm c}$). For $\nu > \max(\nu_{\rm m},\nu_{\rm c})$, 
we have $-p/2$. Here, $p$ is the spectral index of the emitting 
electrons: $N(E)\propto E^{-p}$.

\section{Dynamical Evolution of the Fireball}

During the $\gamma$-ray bursting phase and the early stage of afterglows, the fireball expansion is initially ultra-relativistic and highly 
radiative, but finally it would be getting into non-relativistic and adiabatic, 
a unified dynamical evolution should match all these phases. In fact, the 
initial ultra-relativistic phase has been well described by some simple scaling 
laws (M\'{e}sz\'{a}ros \& Rees 1997a; Vietri 1997; Waxman 1997a; Wijers et al. 1997), while the final 
non-relativistic and adiabatic phase should obey the Sedov (1969) rule, 
which has well been studied in Newtonian approximation. The key equation
(Blandford \& McKee 1976; Chiang \& Dermer 1999) is
\begin{equation}
\frac{{\rm d}\gamma}{{\rm d}m} = -\frac{\gamma^{2}-1}{M},
\end{equation}
here $m$ denotes the rest mass of the swept-up medium, $\gamma$ the bulk Lorentz 
factor, and $M$ the total mass in the co-moving frame including internal energy $U$. 
This equation was originally derived under the ultra-relativistic condition. The 
widely accepted results derived under this equation are correct for 
ultra-relativistic expansion. Accidentally, these results are also suitable for the 
non-relativistic and radiative case. However, for the non-relativistic and 
adiabatic case, they will lead to wrong result ``$v \propto R^{-3}$'' ($v$ is the 
velocity), while the correct Sedov result should be ``$v \propto R^{-3/2}$'', 
as first pointed out by Huang, Dai \& Lu (1999a,b).

It has been proved (Huang, Dai \& Lu 1999a,b) that in the general case, 
the above equation should be replaced by 
\begin{equation}
\frac{{\rm d}\gamma}{{\rm d}m} = -\frac{\gamma^{2}-1}{M_{ej} + \epsilon m 
+ 2(1-\epsilon)\gamma m},
\end{equation}
here $M_{ej}$ is the mass ejected from GRB central engine, $\epsilon$ is the 
radiated fraction of the shock generated thermal energy in the co-moving 
frame. The above equation will lead to correct results for all cases 
including the Sedov limit. This generic model is suitable for both 
ultra-relativistic and non-relativistic, and both radiative and adiabatic 
fireballs. As proved by Huang et al. (1998a,b), Wei \& Lu (1998a) and 
Dai et al. (1999a), 
only several days after the burst, a fireball will usually become 
non-relativistic and adiabatic, while the afterglows can last observable for some months, 
the above generic model is really useful and important.

\section{Comparison and Association of GRB with SN}

Supernova was known as the most energetic phenomenon at the stellar level. 
SN explosion is the final violent event in the stellar evolution. 
Dynamically, it can also be described as a fireball, which however expands 
non-relativistically. After the SN explosion, there is usually a remnant which 
can shine for more than thousands of years and be well described dynamically by 
Sedov model (Sedov 1969).

GRB is also a phenomenon at the stellar level. However, it is much more 
energetic and much more violent than SN explosion! It has been proved to 
be described 
as a fireball, which expands ultra-relativistically. The GRB may also leave
a remnant which shines for months now known as afterglow. 

Their comparison is given in Table I:
\begin{center}
\begin{tabular}{lll}
\multicolumn{3}{c}{Table I} \\ \hline
      &  GRBs    &    SNe    \\   \hline
  {\bf Burst}  &  {\bf Bursting $\gamma$-rays}  &   {\bf SN explosion}  \\ \hline   
  Energy up to     &  10$^{54}$ ergs &   $10^{51}$ ergs   \\
  Time Scale &  10 sec &   Months   \\
  Profile  &  irregular   &   smooth \\
  Wave Band  &  $\gamma$-ray  &  Optical \\  \hline
{\bf Relic}  &  {\bf Afterglow}  &  {\bf Remnant} \\ \hline
Time Scale  &  Months  &  10$^{3}$ Years \\
Wave Band  &  Multi-band  &  Multi-band  \\ \hline
{\bf Understanding}  &  &  \\  \hline
Fireball Expansion  &  Ultra-relativistic  &  Non-relativistic  \\
Mechanism  &  ???  &  Stellar Core Collapse  \\
Key Process  &  ???   &   Neutrino process  \\  \hline
\end{tabular}
\end{center}

In April 1998, a SN 1998bw was found to be in the 8' error circle of the X-ray
afterglow of GRB 980425 (Galama et al. 1998; Kulkarni et al. 1998). However, 
its host galaxy is at a red-shift z=0.0085 (Tinney et al. 1998), indicating a distance of 38 Mpc (for $H_{0} = 65$ km s$^{-1}$ Mpc$^{-1}$), which leads the 
energy of the GRB to be too low, only about 
$5 \times 10^{47}$ ergs, 4 orders of magnitude lower than normal GRB. 

Later, in the light curves of GRB 980326 (Bloom et al. 1999; 
Castro-Tirado \& Gorosabel 1999b) and GRB 970228 (Reichart 1999; 
Galama et al. 2000), some evidence related with SN was found. This is 
a very important question worth while to study further (see e.g. Wheeler 
1999). These two violent phenomena, GRB and SN, might be closely related. 
They might be just two steps of one single event (Woosley et al. 1999; 
Cheng \& Dai 1999; Wang et al. 2000b; Dai 1999d). It is interesting to 
note that the first step might provide a low baryon environment for 
the second step to produce GRB. Such a kind of models can give a way 
to avoid the baryon contamination.

\section{Post-Standard Effects}

The standard model described above is based on the following simple assumptions: 
(1) the fireball expanding relativistically and isotropically; (2) impulsive 
injection of energy from inner engine to the fireball(s); (3) synchrotron radiation 
as the main radiation mechanism; (4) uniform environment with typical particle number density 
of $n=1$ cm$^{-3}$. This model is rather successful in that its physical 
picture is clear, results obtained are simple, and
observations on GRB afterglows support it at least qualitatively but generally. However, 
various quantitative deviations have been found. Thus, the simplifications made in the 
standard model should be improved. These deviations may reveal important new information. 
In the following, we will discuss some of the effects due to these 
deviations (the post-standard effects) (Lu 1999, 2000; Dai 2000e).

\hspace*{1cm} {\bf (i)Environment effects:} 

In the early days after the discovery of afterglows, Dai and Lu(1998c) studied the 
possible non-uniformity of the surrounding medium. They used the general form of 
$n \propto R^{-k}$ to describe the non-uniform environment number density. By fitting 
the X-ray afterglow of GRB 970616, they found $k=2$ which is just the form of a wind 
environment. This indicates that the surrounding medium of GRB 970616 was 
just a stellar wind. After the detailed studies by Chevalier 
and Li (1999, 2000), the stellar wind model for the environment of
GRBs has now become widely interested. As the properties of GRBs' environment contain 
important information related with their pregenitors, this stellar wind model 
provides strong support to the view of massive star origin of GRBs.

Another environment effect is due to the deviation from the standard number density of 
$n=1$ cm$^{-3}$. Some afterglows of GRBs show that their lightcurves obey a broken power law. 
For example, according to Fruchter et al.(1999), the optical lightcurve of GRB 990123 
shows a break after about two days, its slope being steepened from $-$1.09 to $-$1.8. Dai and 
Lu (1999b) pointed out that a shock undergoing the transition from a relativistic phase to 
a non-relativistic phase may show such a break in the light curve. If there are 
dense media and/or clouds in the way, this break may happen earlier to fit the observed 
steepening. Recently, Wang, Dai \& Lu (2000a) proved that the dense environment model can also 
explain well the radio afterglow of GRB 980519 (Frail et al. 2000).

\hspace*{1cm} {\bf (ii)Additional energy injection:} 

Lightcurves of some optical afterglows even show the 
down-up-down variation such as GRB 970228 and 970508. These features can be 
explained by additional long time scale energy injection from 
their central engines (Dai \& Lu 1998a, b; Rees \& M\'{e}sz\'{a}ros 1998; Panaitescu et al. 1998).
In some models,  a millisecond pulsar with strong magnetic field can be produced
at birth of a GRB. As the fireball expands, the central pulsar can continuously
supply energy through magnetic dipole radiation. Initially, the energy supply is
rather small, the afterglow shows declining. As it becomes important, the afterglow
shows rising. However, the magnetic dipole radiation should itself attenuate later.
Thus, the down-up-down shape would appear naturally. Dai \& Lu (2000a) further analysed
GRB 980519, 990510 and 980326, with dense environment also being taken into
account, and found results agreeing well with observations. Wang \& Dai (2000e)
further considered both homogeneous and wind external media, the R-band light curve of GRB 000301c 
was also well fitted.

Recently, the GRB 000301c afterglow shows three break
appearance in the R-band light curve, and extremely steep decay slope $-3.0$ 
at late time. This unusual afterglow can be explained by assuming more 
complicated additional energy injections and dense medium (Dai \& Lu 2000b).

\hspace*{1cm} {\bf (iii)Additional radiations:} 

Though synchrotron radiation is usually thought to be the main radiation mechanism, 
however, under some circumstances, the inverse Compton scattering may play an 
important role in the emission spectrum, and this may influence the temporal properties 
of GRB afterglows (Wei \& Lu 1998a, b, 2000a). Wang, Dai \& Lu (2000c)
even consider the inverse Compton
scattering of the synchrotron photons from relativistic electrons in the
reverse shock. Under appropriate physical parameters of the
GRBs and the interstellar medium, this mechanism can excellently account for
the prompt high energy gamma-rays detected by EGRET, such as from GRB 930131.
It is interesting to note that during the GRB phase, not only electromagnetic radiations, but also
neutrinos, will be emitted (Waxman \& Bahcall 1997, 2000; Halzen 1998; Dai \& Lu 2000c).
The prompt neutrino emission from reverse shocks as a result of the interaction of relativistic 
fireballs with their surrounding wind pointed out by Dai \& Lu (2000c) may be very important.

Later data in the afterglows of GRB 970228 (Reichart 1999; 
Galama et al. 2000) and 980326 (Bloom et al. 1999; Castro-Tirado \& 
Gorosabel 1999b) may show the deviations as additional contributions
from supernovae.

\hspace*{1cm} {\bf (iv)Beaming effects:} 

GRB 990123 has been found very strong in its 
$\gamma$-ray emission, and the red 
shift of its host galaxy is very large (z=1.6) (Kulkarni et al. 1999a; 
Galama et al. 1999; Akerlof et al. 1999; Castro-Tirado, et al. 1999a; 
Hjorth, et al. 1999; Andersen, et al. 1999). If its radiation is isotropic, 
the radiation energy only in $\gamma$-rays is already as high as  
$E_{\gamma} \sim 3.4 \times 10^{54}$ ergs, closely equals two solar rest 
energy ($E_{\gamma} \approx 2{\rm M}_{\odot}$c$^{2}$)! As the typical mass of the 
stellar object is in the order of $\sim 1 {\rm M}_{\odot}$, while the radiation 
efficiency for the total energy converting into the $\gamma$-ray emission 
is usually very low, such a high emission energy is very difficult to 
understand (Wang, Dai \& Lu 2000d).

As some GRBs showed their isotropic radiation energy to be as high as 
$\sim $M$_{\odot}$c$^2$, this has been regarded as an energy crisis.
A natural way to relax this crisis is to assume that the radiation of GRB 
is jet-like, rather than isotropic. Denote the jet angle as $\Omega$, 
then the radiation energy $E$ will be reduced to $E\Omega/4\pi$. At the same time, 
the estimated burst rate should increase by a factor of $4\pi/\Omega$. 
However, we should find out its observational 
evidences. Rhoads (1997, 1999) analysed this question, and predicted that the 
sideways expansion in jet-like case will produce a sharp break in the GRB afterglow
light curves (see also Pugliese et al. (2000), Sari et al. (1999) and 
Wei \& Lu (2000b, c) ). Kulkarni et al. (1999c)
regarded the break in the light curve of GRB 990123 as the evidence for jet. 
However, Panaitescu \& M\'{e}sz\'{a}ros (1999), Moderski, Sikora \& Bulik (2000)
performed numerical calculation and denied the appearence of such a sharp break.
Wei \& Lu (2000b) re-analysed 
the dynamical evolution of the jet blast wave and found that a sharp break 
can only exist in the case of extremely small beaming angle. Huang et al. (2000a, b) 
made a detailed calculation and proved that the breaks in the lightcurves are 
mainly due to the relativistic to non-relativistic transition, not due to edge effect and lateral
expansion effect of the jet, and may appear only
for small electron energy fraction and small magnetic energy fraction. However, they stressed that 
the afterglows of jetted ejecta can be clearly characterized by rapid fading 
in the non-relativistic phase with index $\alpha \ge 2.1$ (Huang et al 2000c).
Recently, Dai, Huang and Lu (2000f) pointed out that the effect of dust extinction 
on jetted GRB afterglows may obviously enhance the breaak in their light curves.

Gou et al. (2000) used a set of refined dynamical equations and a realistic lateral 
speed of the jet, calculated the evolution of a highly collimated jet that expands in a 
stellar wind environment and the expected afterglow from such a jet. They found that 
in the wind environment, no obvious break will appear even at the time when the 
blast wave transits from the relativistic phase to the non-relativistic phase, and 
there will be no flattening tendency even up to $10^{9}$ s. Further calculations on the 
anisotropic jets expanding in different kinds of wind have also been made (Dai \& Gou 2000d). 
GRB 991208 was argued to arise from a highly anisotropic jet expanding in the wind of a red supergiant.

\section{Central Engines and Energy Sources}

It is commonly agreed that the nature of the central engine, which powers the fireball into space, creates shocks where burst of gamma-ray emission and the associated afterglow are emitted, is the most difficult part of GRB-modelling. 
In fact there is no consensus on the GRB central engine even after the discovery of the afterglow. The main reason is that most observed properties of GRBs are emitted from regions far away from the central engine. Since there are less constraints in the central engine models, it is not surprised to find that there are over hundred proposed central engine models in the literatures. Here we can only introduce popular models and some selected possible models. 

\subsection{Popular Cosmological Central Engine Models}

Although there are no compelling direct observed facts to pin-point what are the central engine, there are number of evidence including energy budget, beaming, host galaxies, association with the star formation region (Hogg \& Fruchter 1999) and association with the type Ic supernova (Galama et al. 1998) to constraint the possible central engine models. The following models are widely quoted in the GRB papers:

\hspace*{1cm} {\bf (i)Merger of Two Compact Objects}. 

These two compact object systems could be neutron star-neutron star, neutron star-black hole, neutron star-white dwarf.  Binary neutron star mergers are believed to be the most possible merger model for GRB (e.g. Paczy\'nski 1986; Eichler et al 1989;
Dermer 1992; Narayan, Piran and Shemi 1991; Mao and Paczy\'nski 1992; Davies et al. 1994; Ruffert \& Janka 1999). There are three such binary systems observed in our galaxy (Manchester 1999) and there should be at least 30 because of the  beaming effect. Their orbital periods are decreasing  and should merge in the time scale of $10^8$ years (e.g. Taylor \& Weisberg 1989). This gives the merger rate about $10^6 - 10^7$ years per galaxy and consistent with the observed GRB rate (Narayan, Piran \& Shemi 1991). It has been suggested that neutron star-black hole is the more common binary system than neutron star-neutron star system, therefore GRBs result from the merger of these systems (Bethe \& Brown 1998). However, none of such system has been found in our galaxy. After merger, a 2 --- 4 $M_{\odot}$ black hole surrounded by a 0.1 --- 0.2 $M_{\odot}$ thick accretion disk is formed. The black hole can accrete 0.1 $M_{\odot}$ from the disk in a time scale of 10 --- 100 seconds and the gravitational energy of the accreted matter can provide the energy of GRB process. However, GRB 990123 (Kulkarni et al. 1999a; Andersen et al. 1999) requires $E_{\rm iso}$ must be up to a few times 
$10^{55}$ergs for isotropic emission. This causes problem for the merger models. If anisotropic emission
with a beaming factor of $\Delta\Omega/4\pi\sim 0.01$ 
is assumed, this energy can be reduced 
to $E_{\rm jet}\sim$ a few times $10^{53}\,$ergs. 

\hspace*{1cm} {\bf (ii)Failed Supernova or Collapsar of Hypernova}

 These models propose that massive stars can collapse to form a rapidly spinning black hole surrounded by a thick torus (Woosley 1993; Paczy\'nski 1998). Instead of using the gravitational energy of the accreted matter, these model suggest that the main energy reservoir is the rotation energy of the black hole, which can be extracted via the Blandford-Znajek mechanism (Blandford \& Znajek 1977). In these models, the accretion rate is at least over million time of Eddington rate and the disk magnetic field must be over 10$^{15}$ Gauss. These models are naturally related to supernova, star formation regions, where massive stars are commonly found, and have enough energy budget for the GRB. On the other hand, these central engines are located in the environment filled with baryonic matter. How can a relativistic fireball be developed? It appears to be a problem. Perhaps, a two-step mechanism can resolve this problem. 
Cheng and Dai (2000) have proposed a two-step model for GRBs 
associated with supernovae. In the first step, the core collapse of a star with 
mass $\ge 19M_\odot$ leads to a massive neutron star and a normal supernova, 
and subsequently hypercritical accretion of the neutron star from 
the supernova ejecta may give rise to a jet through neutrino 
annihilation and Poynting flux along the stellar
rotation axis. However, because of too much surrounding matter, this
jet rapidly enters a non-relativistic phase and evolves to a large
bubble. In the second step, the neutron star promptly implodes to
a rapidly rotating black hole surrounded by a torus once the mass of the
star increases to the maximum mass, and meanwhile its rotation frequency
increases to the upper limit due to the accreted angular momentum.
The gravitational binding energy of the torus may be dissipated by
a magnetized relativistic wind, which may then be absorbed by the
supernova ejecta, thus producing an energetic hypernova. The rotational
energy of the black hole may be extracted by the Blandford-Znajek
mechanism, leading to another jet. This jet is relatively free of
baryons and thus may be accelerated to an ultrarelativistic phase
because the first jet has pushed out of its front matter and left
a baryon-free exit. Therefore the second jet generates a long/hard 
GRB and its afterglow. This is  because the energy release timescale 
in the second step is about $E_{\rm rot}/P_{\rm BZ}\sim 10^2$ s, which 
corresponds to the GRB duration,  if the involved magnetic field 
has a strength of the order of $10^{16}$ G, and also because the Lorentz 
factor of the second jet in this model may be larger than 100 so that photons
radiated from this jet can be blue-shifted to hard ones. 
Recently, some authors (Cen 1998; Wang \& Wheeler 1998) envisioned,
in a small cone around some special axis of a newborn neutron star, 
the matter is assumed to be preferentially first blown out in order 
to avoid too many baryons contaminating a subsequently resulting jet. 
Our model may provide a plausible way of how such an empty cone is
produced: neutrinos from the hypercritical accretion disk annihilate 
to electron/positron pairs which form the first jet to push its front
baryons and leave an exit for the second jet. Therefore, this model can
avoid the baryon contamination problem.

\subsection{Possible Cosmological Central Engine Models}

\hspace*{1cm} {\bf (i)Rapidly Spinning and Strongly Magnetized Compact Objects}

 The required energy budget of a typical GRB is $\sim 10^{53}$ ergs,
 which is about the rotational energy of a neutron star rotating in its
 maximum Keplerian speed. The main problem is how to release this amount
 of rotation energy in a time scale of 10 --- 100 seconds, which can be
 achieved if the star has a super-strong magnetic field
 ($\ge 10^{15}$ Gauss). High potential drop ($\sim 10^{20}$ Volts) can
 easily produce a relativistic pulsar wind and subsequent radiation can
 produce the observed GRB spectral properties (Usov 1992; Duncan \&
 Thompson 1992). The recent discovery of Anomaly X-ray Pulsars and
 Soft Gamma-ray Repeaters, which are found to be neutron stars with
 magnetic field strength from $10^{13}-10^{15}$ Gauss, provide support
 for these models. However, these models exist number of disadvantages
 because the gamma-rays are expected to be emitted in the internal
 shocks, which requires that the energy release cannot be continuous
 and some baryons ($\le 10^{-5}M_{\odot}$) must be there. In fact, these requirements are also important to explain the time variability observed in GRBs. It has been suggested that the super-strong magnetic field is produced after the neutron star is formed. The possible magnetic field production dynamo results from the differential rotation of different parts of the star. If this is the case, an even stronger toroidal field $\sim 10^{17}$ G can be produced by differential rotation. When the toroidal field generated inside the star becomes strong enough to break the crust, the emerging field together with the broken platelets can be reconnected on the surface of neutron stars and transfer the magnetic energy to baryonic matter in the broken platelets. This process can continue until all rotation energy is converted into magnetic energy, which eventually goes to the fireball( Klu\'zniak \& Ruderman 1998; Ruderman, Tao \& Klu\'zniak 2000). It was immediately realized that such process could also work for rotating strange stars (Dai \& Lu 1998), who pointed out that the rise of the afterglow in GRB 970508 could result from the continuous energy supply of the central pulsar via pulsar wind. 

\hspace*{1cm} {\bf (ii)Phase Transition of Compact Objects}

 Pion condensation in the core of neutron star has been proposed to be the possible mechanism as the energy source of GRBs (e.g. Ramaty et al. 1980). However, it has been shown that such process cannot be a sudden process due to the conservation of charge and baryon number (Glendenning 1992). On the other hand, when a neutron star in Low Mass X-ray Binary accretes over about half a solar mass from its companion, it can undergo a phase transition from neutron star to strange star and release even larger amount of energy 
to account for GRB.  20 --- 30 MeV is released per baryon during the phase transition. Total energy released 
this way can be up to about (4 --- 6)$ \times 10^{52}$ ergs. Strange star is the 
stellar object in the quark level. Whether it exists or not is a fundamental 
physical/astrophysical problem. Its main part is a quark core with large 
strangeness (known as strange core). There could be a thin crust with mass of 
only about $\sim (10^{-6} - 10^{-5})$M$_{\odot}$ (Alcock et al. 1986; Huang \& Lu 1997a,b; Lu 1997; Cheng et al. 1998), all baryons are contained in the crust. 
The resulting strange star has a thin crust with mass $\sim 2\times
10^{-5}M_{\odot}$ and thickness $\sim 150$m, but because the
internal temperature is so high ($\sim 10^{11}$K), the nuclei
in this crust may decompose into nucleons.
Approximating strange matter by a free Fermi 
gas, the total thermal energy of the star is given by
$E_{{\rm th}}\sim 5\times 10^{51}\,{\rm ergs}\,(\rho/\rho_{0})^{2/3}
R_{6}^{3}T_{11}^{2}\,$, where $\rho$ is the average mass density, 
$R_{6}$ the stellar radius in units of $10^{6}$cm,
and $T_{11}$ the temperature in units of $10^{11}$K.
Adopting $\rho=8\rho_{0}$, $R_{6}=1$, and $T_{11}=1.5$, we have
$E_{{\rm th}}\sim 5\times 10^{52}\,{\rm ergs}$.
The star will cool by the emission of neutrinos and antineutrinos, and
because of the huge neutrino number density, the neutrino
pair annihilation process $\nu \bar{\nu}\rightarrow e^{+}e^{-}$ operates in 
the region close to the strange star surface.
The total energy (Haensel, Paczy\'{n}ski \& Amsterdamski 1992) deposited due to this process is
$E_{1}\sim 2\times 10^{48}\,\,{\rm ergs}\, (T_{0}/10^{11}{\rm K})^{4} 
\sim 10^{49}\,{\rm ergs}$ (where $T_{0}$ is the initial temperature) 
and the timescale for deposition is of the order of 1s.
On the other hand, the processes for $n+\nu_{e}\rightarrow p+e^{-}$ and
$p+\bar{\nu_{e}}\rightarrow n+e^{+}$ play an important role in the
energy deposition and the integrated neutrino
optical depth (M\'{e}sz\'{a}ros \& Rees 1992) due to these processes is
$\tau \sim 4.5 \times 10^{-2}\rho_{11}^{4/3}T_{11}^{2}$
(where $\rho_{11}$ is the crust density in units of $10^{11}{\rm g}\,{\rm
cm}^{-3}$). So the deposition energy is estimated by
$E_{2}\sim E_{{\rm th}}(1-e^{-\tau}) \sim 2\times 10^{52}\,{\rm ergs}$.
Here the neutron-drip density ($\rho_{11} \sim 4.3$) is used.
The process, $\gamma \gamma\leftrightarrow e^{+}e^{-}$, 
inevitably leads to creation of a fireball.
However the fireball must be contaminated by the baryons in
the thin crust of the strange star. If we define
$\eta=E_{0}/M_{0}c^{2}$, where $E_{0}=E_{1}+E_{2}$ is the initial radiation 
energy produced ($e^{+}e^{-}$, $\gamma$) and
$M_{0}$ is the conserved rest mass of baryons with which the fireball is
loaded, then, since the amount of the baryons contaminating the fireball cannot
exceed the mass of the thin crust, we have $\eta\ge 5\times 10^{3}$ and
the fireball will expand outward. The expanding
shell (having a relativistic factor $\Gamma \sim \eta$) interacts with
the surrounding interstellar medium and its kinetic energy is finally
radiated through non-thermal processes in shocks.  

Another possible phase transition model is an accreting strange star collapsing into a naked singularity (Harko \& Cheng 2000).

\hspace*{1cm} {\bf (iii)Accretion Onto Massive Black Holes}

 There are central engine models based on massive black holes associated with Quasars or AGNs in galactic centers (e.g. Carter 1992). These models are ruled out as all GRBs with optical afterglow are not associated with such objects (Piran 1999). 
Perhaps this conclusion should be revised as all long duration GRBs are not associated with either Quasars or AGNs because all GRBs with afterglow are long duration bursts. Statistical analysis of GRBs in the BASTE catalog does show some correlation with certain type of Quasars. For example, Schartel, Andernach \& Greiner (1997) found a surprising evidence of a positional correlation between 
GRBs and radio-quiet quasars, but for most classes of 
AGNs, BL Lac objects and radio-loud quasars, no excess 
coincidences above random expectation; 
Marani et al (1997) reported a suggestive correlation 
between GRBs and ACO clusters though some of the 
significance of correlation can be rather high; 
Kolatt \& Piran (1996), Hurley et al (1997), and 
Gorosabel \& Castro-Tirado (1997) studied the 
angular correlation between GRBs and Abell 
clusters but obtained different results.  
There are also many works focused on the studies of 
GRBs repetition (e.g., Quashnock \& Lamb 1993; 
Brainerd et al 1995; Meegan et al 1995; 
Bennett \& Sun 1996) with various statistical methods.  
But the sound of negative results seems loud.

Recently metal rich Quasars are found to correlate with GRBs in 99\% significant level (Cheng et al. 1997).
It can be understood why metal rich Quasars could be the host galaxies of GRBs in the following reasons (Cheng \& Wang 1999). The production rate of compact objects, i.e. neutron stars (NS) 
and black holes (BH), in active galactic nuclei(AGN) and quasars(QSO), where the frequent supernova explosion is used to explain the high metallicity (Artymowicz, Lin \& Wampler 1993; Zurek et al 1994), is very high due to the interaction between the accretion disk and main sequence
stars in the nucleus of the quasar. The compact object-red giant star(RG) binaries can be easily formed due to the large captured cross-section of the red giant stars. 
The (NS/BH, NS/BH) binary can be formed after the supernova
explosion of the (NS/BH, RG) binary. Intense transient gamma-ray emission (gamma-ray burst) and gravitational radiation can result from the merger of these two compact objects. Collision between helium core (Hc) of RG and black hole may also take place and  may also result in long duration gamma-ray bursts but no gravitational waves (Popham, Woosley \& Fryer 1999). The estimated merger rate of (NS/BH, NS/BH) binaries and (Hc, BH) is proportional to the metal abundance ($\frac{NV}{CIV}$) and can be as high as 10$^{-3}(\frac{NV}{CIV}/0.01)$ per year per AGN/QSO.

Massive black hole is not favorable central engine because of the short time variability, which suggests that the central engine should be a stellar mass compact object. However, this is misleading because the relativistic motion of the fireball can reduce the time scale in the GRBs (Lu, Cheng, Zhao \& Yang 2000). Cheng and Lu (2001) also suggest that an extreme Kerr black hole with a mass $\sim 10^6M_\odot$, a dimensionless angular momentum $A\sim 1$ and 
 a marginally stable orbital radius $r_{\rm ms}\sim 3r_{\rm s}\sim 10^{12}M_6~$cm
 located in a normal galaxy, may produce a Gamma-ray Burst by capturing and
disrupting a star.  During the capture period, a transient accretion disk is formed
and a strong transient magnetic field $\sim 2.4\times 10^9M_6^{-1/2}$
G (Haswell et al. 1992),
 lasting for $r_{\rm ms}/c\sim 30 M_6~$s,
 may be produced at the inner boundary of the accretion disk.
A large amount of rotational energy of the black hole is extracted
and released in an ultra relativistic jet with a bulk Lorentz factor $\Gamma$ larger
than $10^3$ via the Blandford-Znajek process.
 The relativistic jet energy can be converted into $\gamma$-ray radiation
 via an internal shock mechanism.
The GRB duration should be the same as the life
 time of the strong transient magnetic field. The maximum number of
 sub-bursts is estimated to be $r_{\rm ms}/h\sim (10 - 10^2)$ because the disk material is 
likely to break into pieces with a size about the thickness of the disk 
$h$ at the cusp ($2r_{\rm s}\le r \le 3r_{\rm s}$). The shortest rise time of the
burst estimated from this model is 
$\sim h/\Gamma c\sim 3\times 10^{-4}\Gamma^{-1}_3(h/r)_{-2}M_6$ s. The model gamma-ray burst density rate is also estimated to be consistent with the observed GRB burst rate. Furthermore, the optical afterglow can also be understood in this model. Because in the early stage of forming transient accretion disk, the material
supplied by the tidally disrupted star is clumped in the relatively cool disk.
This eventually leads to a thermal instability, which results in an
increase of the viscosity of the disk. 
As the instability propagates across the disk, the stability of a
time-dependent disk and variability of the mass-deposition rate 
provide a possible explanation for 
 the behavior of GRB optical afterglow (Lu, Cheng \& Zhao 2000).

\subsection{Galactic Central Engine Models}

Although the cosmological origin of GRBs are strongly 
supported by the observation, at least a fraction of GRBs
originating from the Galactic neutron stars is also suggested
by the observation of absorption lines and soft X-ray emission 
from some bursts (but BATSE has not seen such line features yet). 
About $20\%$ of GRBs in the KONUS catalog show
single absorption features at energies between about 20 and 60
keV (Mazets et al. 1981) and three bursts seen by Ginga detector
show double absorption features in the same energy range (Murakami
et al. 1988). These features are usually attributed to the
cyclotron resonant scattering of photons in strong magnetic
fields of the order of $\sim 10^{12} G$ near the neutron star surface.
Therefore it is reasonable to think that at least a subclass of
GRBs may be originated in our Galaxy. The 
Galactic models also result in an isotropic distribution because the bursts are produced by
high velocity neutron stars born in the vicinity of the disk flowing
out into the halo (e.g. Li \& Dermer 1992; Podsiadlowski, Rees \& Ruderman
1995; Wei \& Lu 1996). Observations show that the mean 
pulsar birth velocity is about 450 km s$^{-1}$ (Lyne \& Lorimer
1994), which provides some support to such models. The energy sources basically are accretion energy, thermonuclear flash, rotation energy and internal energy of neutron stars (for a general review of galactic GRB models cf. Harding 1991). But the most promising models are starquake models. As a neutron star quakes, the oscillation of the magnetic field lines
anchored in the crustal platelet produces Alfven wave. 
Blaes et al. (1989) have argued that such Alfven wave can accelerate charged particles and produce GRBs. 
However, the question is can these high velocity neutron stars constitute
a subclass of observed GRBs? 
Hartmann \& Narayan (1996) considered the global energy requirements 
of such models and argued that the rotational energy of
neutron stars is insufficient by, at least, orders of magnitude to provide
the observed burst rate because the origin of the starquake energy comes from the crustal superfluid, which is only 1\% of the stellar rotation energy. Wei and Cheng (1997) also carried a statistical analysis by using the observed period distribution of pulsars and concluded the same result. On the other hand it is possible to make use of full rotational energy, if pulsar activities can be re-ignited in these old neutron stars (Ruderman \& Cheng 1988; Cheng \& Ding
1993; Ding \& Cheng 1997). It is suggested that the detail study of GRB spectral features at lower and higher energy components can be a possible way to distinguish between cosmological and Galactic origin of GRBs (Wei \& Cheng 1997).

\section{Outlook}

Indeed the discovery of the red-shifts of the host galaxies of GRBs in afterglows is a major breakthrough in understanding the nature of GRBs (van Paradijs et al. 2000). The subsequent study of spectral evolution of afterglows in terms of fireball models reveals many important features of the environment of GRBs.
The standard internal-external shock model, which is built under many simplifications, has been proved to be well fitted by 
observations qualitatively but generally. Based on the success of this 
model, it should be very important to study the deviations from the 
standard model, which indicate that the simplifications should be 
relaxed in some aspects. Hence, the deviations contain 
important new information and have been a fruitful research area. 

In contrast to the rapid progress in understanding the nature of 
afterglows, GRB itself has not yet been clear, i.e. the central engine. This is a very important problem. In fact, a viable model for the GRB central engine must address to the following problems:

\hspace*{1cm} {\bf (i)Energy Budget}

The central engine must be able to release more than $10^{53}$ ergs or even $10^{54}$ ergs to radiation. The beaming factor can reduce the energy budget by a factor of 10-100, however, the conversion efficiency from particle kinetic energy to radiation energy likely increases the energy budget by a factor of 10. In fact, most merger models cannot release that much energy (Suen 2000, private communication).

\hspace*{1cm} {\bf (ii)Baryon Contamination}

 Another problem for existing central engine models is the baryon contamination. In order to explain the short time scales of the GRBs, the fireball must be relativistic. This requires the baryonic mass in the fireball must be less than $10^{-5} (E/2 \times 10^{51}{\rm ergs}) M_{\odot}$. The failed supernova (e.g. Woosley 1993), collapsar of hypernova (e.g. Paczynski 1998), newly formed rapidly spinning and superstrong magnetic field neutron stars (e.g. Usov 1992; Kluzniak \& Ruderman 1998; Ruderman, Tao \& Kluzniak 2000) suffer from this problem. Models without baryon contamination problem, e.g. conversion of neutron star to strange star by accretion (Cheng \& Dai 1996; Dai \& Lu 1998), cannot explain GRB-SN association (Galama et al. 1998; Kulkarni et al. 1998). In fact, this is a dilemma of GRB-SN association unless some mechanisms, which can reduce the baryon contamination in supernova indeed exist (e.g. a two-step model proposed by Cheng \& Dai 2000). 

\hspace*{1cm} {\bf (iii)Birth Rate}

 It must be larger than $10^{-6}$ per galaxy per year. The actual rate should be much larger than this if beaming exists.

\hspace*{1cm} {\bf (iv)Various Time Scales}

 The sub-structure of the light curves of GRBs can be as short as 10$^{-3}$s but the duration of GRBs can be as long as 10$^3$s. The former suggests that the emission must be from very small volume and is normally interpreted as that the central engine must be a solar mass compact object, but the latter indicates that the energy release cannot take place in the gravitational time scale. In fact, there are a great variety of light curves in GRBs, which suggest that the energy release process cannot be a simple one.

\hspace*{1cm} {\bf (v)Acceleration and Microphysics}

 The central engine must efficiently accelerate the ejecta to extremely relativistic. How do the collisionless shocks arise within the emitting region? How do these shocks accelerate the particles and enhance the magnetic fields?

Perhaps it is misleading to use the current observations to limit GRB models, i.e. GRB-SN association, host galaxies etc. because they are all long duration GRBs. It is known that GRB durations scatter over six orders of magnitudes (from 1 ms to 1 ks), but a ``typical'' GRB lasts about 10 s. The distribution of burst durations is bimodal, i.e., GRBs can be divided into two sub-groups: long bursts with $T_{90} > 2$ s and short bursts with $T_{90} < 2$ s (Piran 1999). The ratio of observed long bursts to short bursts is three to one. For example, HETE-II, which was successfully launched recently and should be able to detect short duration bursts, may provide new information to constraint GRB models. In our opinion, we are far from understanding the nature of these mysterious GRB events. More information from future GRB missions in electromagnetic wave bands (for a review cf. Hurley 1999), in gravitational wave bands (e.g. LIGO, VIRGO, TAMMA300 and GEO600, for a review of various gravitational detectors cf. Abramovichi et al. 1992; Bardachia et al. 1990; Throne 1997) and low energy neutrino detection,   which can differentiate various merger models (Clark \& Eardley 1977), should help us to solve the puzzle of these important astrophysical phenomena.

\acknowledgements{ We are happy to thank L.X. Cheng, S. Colgate, Z.G. Dai, K.Y. Ding, T. Harko, Y.F. Huang, Y. Lu, M. Ruderman, W.M. Suen, J.M. Wang, D.M. Wei for their useful conversations in various topics in GRBs. This work is partially supported by a RGC grant of the Hong Kong Government and the National Natural Science Foundation of China.
}

\clearpage

\begin{center}
{\bf{Figure Caption}}
\end{center}

\figcaption{Typical light curves detected by BASTE (Greiner 1999) 
\label{fig_1}}

\figcaption{Spatial Distribution of 1869 GRBs (Greiner 1999) 
\label{fig_2}}

\figcaption{Spectra of two GRBs (Schaefer et al. 1998) 
\label{fig_3}}

\end{document}